# A Reconfigurable Active Huygens' Metalens

*Ke Chen[1], Yijun Feng[1]\*, Francesco Monticone[2], Junming Zhao[1], Bo Zhu[1], Tian Jiang[1], Lei Zhang[3], Yongjune Kim[3], Xumin Ding[4], Shuang Zhang[5], Andrea Alù[2]\*, and Cheng-Wei Qiu[3]\**

Dr. Ke Chen, Prof. Yijun Feng, Dr. Junming Zhao, Dr. Bo Zhu, Dr. Tian Jiang

School of Electronic Science and Engineering, Nanjing University, Nanjing 210093, China.

Dr. Francesco Monticone, Prof. Andrea Alù

Department of Electrical and Computer Engineering, The University of Texas at Austin, 1 University Station C0803, Austin, Texas 78712, USA

Dr. Lei Zhang, Dr. Yongjune Kim, Prof. Cheng-Wei Qiu

Department of Electrical and Computer Engineering, National University of Singapore, Singapore 117583, Singapore.

Dr. Xumin Ding

Department of Microwave Engineering, Harbin Institute of Technology, Harbin 150001, China

Prof. Shuang Zhang

School of Physics & Astronomy, University of Birmingham, Birmingham B15 2TT, UK.

\*Correspondence and requests for materials should be addressed to Y.F. (email: yjfeng@nju.edu.cn), A.A. (email: alu@mail.utexas.edu) or to C.-W.Q. (eleqc@nus.edu.sg).

**Abstract:** Metasurfaces enable a new paradigm of controlling electromagnetic waves by manipulating subwavelength artificial structures within just a fraction of wavelength. Despite the rapid growth, simultaneously achieving low-dimensionality, high transmission efficiency, real-time continuous reconfigurability, and a wide variety of re-programmable functions are still very challenging, forcing researchers to realize just one or few of the aforementioned features in one design. In this study, we report a subwavelength reconfigurable Huygens' metasurface realized by loading it with controllable active elements. Our proposed design provides a unified solution to the aforementioned challenges of real-time local reconfigurability of efficient Huygens' metasurfaces. As one exemplary demonstration, we experimentally realized a reconfigurable metalens at the microwave frequencies which, to our best knowledge, demonstrates for the first time that multiple and complex focal spots can be controlled simultaneously at distinct spatial positions and re-programmable in any desired fashion, with fast response time and high efficiency. The presented active Huygens' metalens may offer unprecedented potentials for real-time, fast, and sophisticated electromagnetic wave manipulation such as dynamic holography, focusing, beam shaping/steering, imaging and active emission control.

Metasurfaces, the two-dimensional (2D) metamaterials, have recently attracted significant attention due to their ability to create arbitrary electromagnetic (EM) wavefronts by introducing corresponding field discontinuities across the interface. Many intriguing EM devices have been implemented based on metasurface techniques, including couplers,[1] beam shapers,[2,3] wave-plates,[4] invisibility cloaks,[5,6] imaging systems,[7-9] and other functional devices.[10-18] Most metasurfaces are composed of passive building blocks, such as structured metallic or dielectric resonators, which lack tunable or reconfigurable response. Therefore, one metasurface only empowers one or a few functions mentioned above, inherently limiting their practical impact. In recent years, more efforts have been devoted to achieve active control of metamaterials and metasurfaces by considering tunable or switchable meta-atoms driven by mechanically actuation,[19,20] thermal effects,[21,22] electric voltage bias,[23-28] etc. Tunable or reconfigurable metamaterials of these kinds have been designed and implemented for different practical applications such as tunable absorbers,[29,30] reconfigurable antennas,[31] EM wave modulators,[32,33] and tunable chiral metamaterials.[34,35] In microwave frequencies, by utilizing active elements to switch the meta-atoms with just few discrete levels of phase responses, e.g., 0 and $\pi$ in most cases, recent works demonstrated that tunable metasurfaces have promising prospects in real-time EM wavefront control.[36-42] Ideally, the response of each meta-atom should be controlled *individually, continuously, and reversibly*, which will greatly improve the degrees of freedom in the design of achievable reconfigurable functions but clearly brings substantial challenges in the design and realization of tunable metasurfaces due to the stringent requirements on continuous phase coverage, impedance matching and high efficiency. It should be distinguished from conventional phased antenna arrays, which can actively shape the radiation patterns by controlling each emitter but with much more complex, bulky and highly expensive phase shifters and feeding networks. For these reasons, it is still rare to simultaneously embrace the low-dimensionality, high transmission efficiency, real-time reconfigurability, and various re-programmable functions in one thin-thickness metasurface.

As fundamental devices for imaging, antenna systems, and many other applications, lenses tailor the EM wavefront through their non-planar geometric profiles. However, the applicability of the conventional lenses is restricted where the available space is limited, for example in nanophotonics and compact imaging systems. Metasurfaces can circumvent these limitations by introducing interfacial phase-discontinuities,[3] supported by tailored surface currents. Numerous flat metalenses and their applications with versatile performances have been recently demonstrated, paving the way to the new paradigm of metasurfaces.[16-18], [43-51] Interestingly, a few groups have also investigated the possibility of designing metalenses being able to control the focal response in real time, which requires the phase behavior of each constituent meta-atom to be dynamically, individually, and reversibly controlled under external stimulations.[52,53] However, the existing methods suffer from various limitations. For example, the liquid-metal-injection method used in ref. 52 to change the structural parameters of the metasurface is not only limited to particular resonator structures but also tunable with a slow response time. The metalens composed of phase-change materials ($Ge_2Sb_2Te_5$) in ref. 53 represents a smart and practical method for individual and reversible control of meta-atoms at optical frequencies; however, its tuning process is rather complicated, as it involves writing, erasing and re-writing the constituent elements by means of laser heating, pixel by pixel. Also, only by rewriting once, the focal behavior degrades significantly as their major figure regarding focusing shows.

In general, several challenges still need to be tackled to realize real-time reconfigurable metasurfaces, based on practically deployable solutions. One of the main challenges associated with any passive or active metasurface operating in transmission is the ability to guarantee high transmission efficiency, while at the same time being able to fully control the transmission phase with subwavelength resolution. This can be achieved by using the so-called Huygens' metasurfaces,[54-60] which provide full 2π of phase coverage with high manipulation efficiency via impedance-matching with respect to free space, by suitably engineering both electric and magnetic polarizabilities of the meta-atoms. Therefore, tunable Huygens' metasurfaces by including active lumped components into the meta-atom design may provide an efficient way to arbitrarily and independently tune the transmission amplitude and phase responses. Here we demonstrate an actively tunable Huygens' metalens by employing voltage-controlled varactors into the resonating meta-atoms, whose individual phase responses can be dynamically and continuously manipulated at will. The proposed design is practically implemented at microwave frequencies, demonstrating its feasibility for real-world applications. Since each element can be individually addressed and tuned dynamically by the bias voltages, such metalens can reshape the incident EM wave into predetermined wavefronts, allowing reconfigurable focusing performance that would be otherwise difficult or impossible with conventional lenses. In particular, we demonstrate that multiple focal spots can be realized simultaneously at distinct spatial positions, and that can be moved along pre-designed trajectories, with high efficiency and fast response time, by simply changing the spatial distribution of control voltage applied to each varactor. Both experiment and simulation match well with each other. Therefore, we believe the proposed thin-thickness, efficient, and active metalens, with simultaneously real-time, spatial reconfigurability, are expected to offer untapped potential for arbitrary dynamical focusing of electromagnetic signals, for dynamic holography, active EM wave emission control, etc.

The proposed active Huygens' metasurface consists of a 2D square array of both electric and magnetic resonant meta-atoms loaded with lumped elements with tunable capacitance, as shown in **Figure 1**a. The total thickness of the metasurface is about $0.16\lambda$ ($\lambda$ is the working wavelength in free space). The inset of Figure 1b shows the designed meta-atom, composed of subwavelength metallic structures etched on a dielectric substrate (with a relative permittivity of 2.6 and a loss tangent of 0.002). The meta-atom can be recognized as an electrically small Huygens' source, with magnetic and electric dipoles supported by two metallic split rings and a pair of cut wires, respectively. When a propagating wave impinges on such Huygens' metasurface, transverse magnetic and electrical currents will be induced (details are shown in Figure S4), which are determined by the electric and magnetic polarizabilities of the meta-atoms, respectively. In addition, if the meta-atoms are subwavelength and densely packed, the response of the metasurface could be effectively described by an averaged surface impedance. Assuming that the transverse variations in the plane of the metasurface are slow, the boundary condition presented to the impinging wave can be locally defined in terms of an averaged electric sheet admittance ($Y_{es}$) and an averaged magnetic sheet impedance ($Z_{ms}$). Then the local complex transmission and reflection coefficients can be obtained by:[54]

$$T = \frac{2}{2+\eta Y_{es}} - \frac{Z_{ms}}{Z_{ms}+2\eta}, \qquad (1)$$

$$R = \frac{-\eta Y_{es}}{2+\eta Y_{es}} + \frac{Z_{ms}}{Z_{ms}+2\eta}, \qquad (2)$$

where $\eta$ is the wave impedance of free-space. By appropriately engineering the meta-atoms an arbitrary local discontinuity, including both local amplitude and phase, can be imparted on the impinging field according to the equivalence principle.[54-60] Therefore, a flexible and efficient method to engineer the wavefront of EM waves could be anticipated.

Here, in order to realize the abovementioned functionalities, we design an active Huygens' metasurface with sheet impedance that can be locally tuned by an external electric signal. In particular, the tuning of the transmission phase can be controlled by loading the meta-atoms with electrically driven varactors. In the present scheme, three varactors are loaded in the meta-atom in order to control such impedance, as further detailed in the Supporting Information (SI). For design and illustration simplicity, the varactor capacitances are assumed to have the same numeric value. The geometric parameters of the structure are optimized to achieve a nearly 360 degrees coverage of available phase shifts when the sheet impedance is tuned by varying the varactor capacitance from 0.05 to 3 pF, while preserving a relative high transmission associated with the Huygens nature of the designed meta-atom, as shown in Figure 1b. Such continuously tunable phase variation can be exploited to implement arbitrary spatial phase gradients across the metasurface with high efficiency in the manipulation of the transmitted beam.

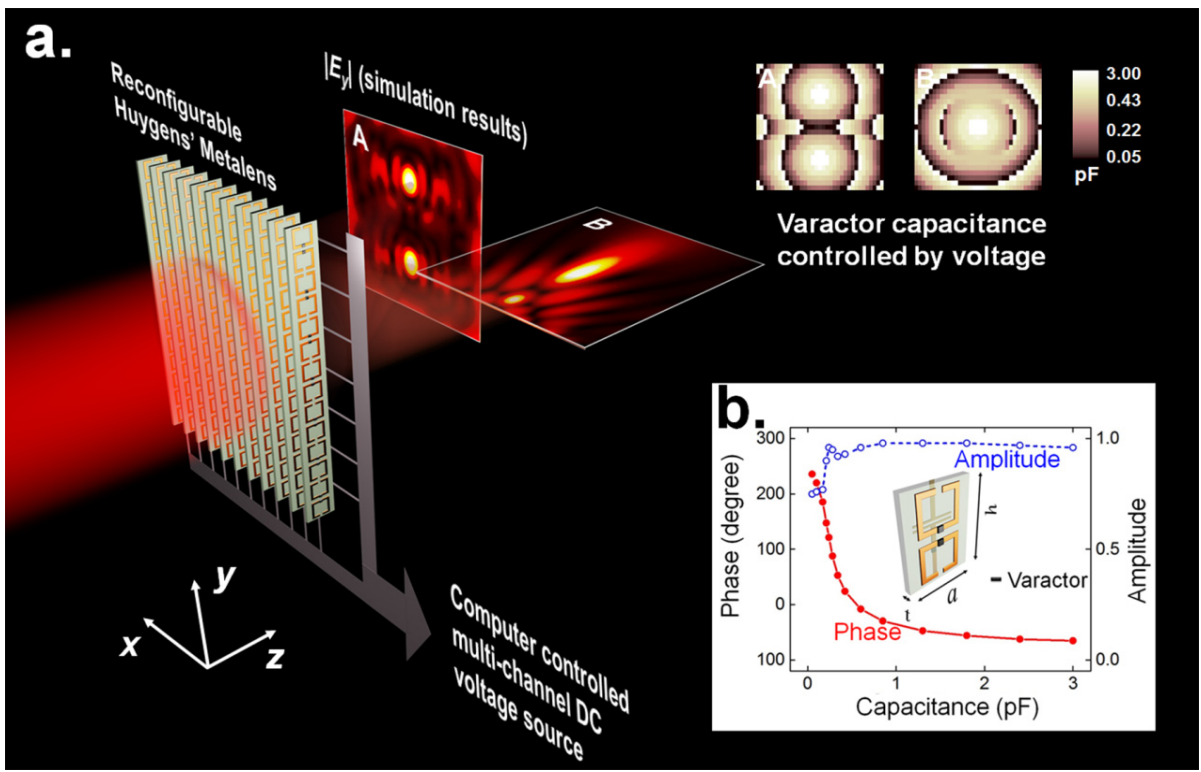

**Figure 1.** Reconfigurable Huygens' metalens and characteristics of meta-atom. **(a)** Active Huygens' metalens for dynamic EM wave focusing, of which meta-atoms are biased by computer-controlled multi-channel DC voltage sources. Upper-right inset: spatial distributions of capacitances used to achieve two focal spots either in *x-y* (A) or in *x-z* plane (B) based on full-wave simulation. **(b)** Capacitance-dependent phase and amplitude responses of EM wave transmission at the target frequency of 6.9 GHz for incident electric and magnetic fields which are parallel to *y*- and *x*-axes, respectively. Inset: schematic of meta-atom with geometric parameters of $a$ = 7 mm, $h$ = 10 mm, $t$ = 0.8 mm, and the periodicity along *x*-direction is 7 mm.

To achieve an arbitrary focal pattern, the impedance distribution across the metalens aperture has to be designed, which transforms the incident plane wave into the desired focused beam. To this end, we analyze the electromagnetic problem similar to the process of holographic imaging.[7] We select the ideal dipolar sources as the virtual sources, and arrange them on the positions of the pre-designed foci. In general, we can assume that there are $N$ focal spots located at $(x_i, y_i, z_i)$ ($i$=1 to $N$). For example, the case A of Figure 1a requires two sources located at $(x_1 = 0, y_1 = 1.5\lambda, z_1 = \lambda)$ and $(x_2 = 0, y_2 = -1.5\lambda, z_2 = \lambda)$, respectively. By assuming in-phase radiation from all virtual sources, the superposition of the radiated fields is mapped onto the recording plane where the metalens is located ($z = 0$), and the interference electric field is retrieved as[61]

$$E_t(x,y,0) = \sum_{i=1}^{i=N} \hat{y} \{ \frac{A_i}{4\pi} j\omega\mu_0 [1+\frac{1}{j\beta r_i}-\frac{1}{(\beta r_i)^2}]\frac{e^{-j\beta r_i}}{r_i}\sin\theta_i \hat{\theta}_i + \frac{A_i}{2\pi}\eta[\frac{1}{r_i}-j\frac{1}{\beta r_i^2}]\frac{e^{-j\beta r_i}}{r_i}\cos\theta_i \hat{r}_i \} \quad (3)$$

where $r_i = \sqrt{(x_i-x)^2+(y_i-y)^2+z_i^2}$. The angle between the position vector $\vec{r}_i$ and the $y$-axis as well as the relative amplitude of the source are denoted by $\theta_i$ as well as $A_i$, respectively. The phase constant, wave impedance, and permeability in free-space are denoted by $\beta$, $\eta$, and $\mu_0$, respectively. Then, the required surface impedance distribution can be retrieved from the fields on the two sides of the interface, which are the incident field of an uniform plane wave and the field calculated by Equation (3) in the regions of $z<0$ and $z>0$ of Figure 1a, respectively. Considering a normally incident plane wave, both of the electric sheet admittance $Y_{es}$ and the magnetic sheet impedance $Z_{ms}$ are purely imaginary and satisfy $Y_{es}\eta = Z_{ms}/\eta$, which ensures not only impedance matching but also a transmission phase distribution as $-\arg[E_t(x,y,0)]$. Then, the calculated continuous impedance distribution is discretized into $35\times25$ pixels which can be realized by changing the varactor capacitances in the corresponding meta-atoms according to Figure 1b. Other alternative methods can also be applied to realize static multi-foci or arbitrary focal pattern.[62,63]

As mentioned in the Introduction, in order to construct a real-time reconfigurable metasurface, each constituent meta-atom should be individually and dynamically controlled with a short response time. As a proof-of-concept example, we show that a planar metalens composed of individually controlled active Huygens' elements can dynamically focus the incident wave to arbitrary focal spots in three-dimensional (3D) space, as schematically shown in Figure 1a. Two focal spots either in the *x-y* (panel A) or *x-z* plane (panel B) can be achieved by tuning the spatially varying sheet impedances on the lens aperture to the predesigned distribution. The ideal impedance distributions necessary to obtain two foci (case A in Figure 1a) are illustrated in the upper row of **Figure 2**a. The discretized distributions of the imaginary part of $Y_{es}$ and $Z_{ms}$ are shown in the lower row of Figure 2a. A direct comparison of the realized distributions of real and imaginary parts of $Y_{es}$ and $Z_{ms}$ on the *y*-axis with the idea counterparts are also plotted in Figure 2b, showing a good agreement. By tuning the capacitance distribution of varactors on the lens aperture, in principle nearly arbitrary focal patterns can be realized, and more complex focal patterns are shown in Figure S1 of the SI. The advantage of the proposed tunable Huygens' metalens is its compelling ability to dynamically generate various focal patterns, by simply reconfiguring the spatial distribution of the varactor capacitance in each meta-atom.

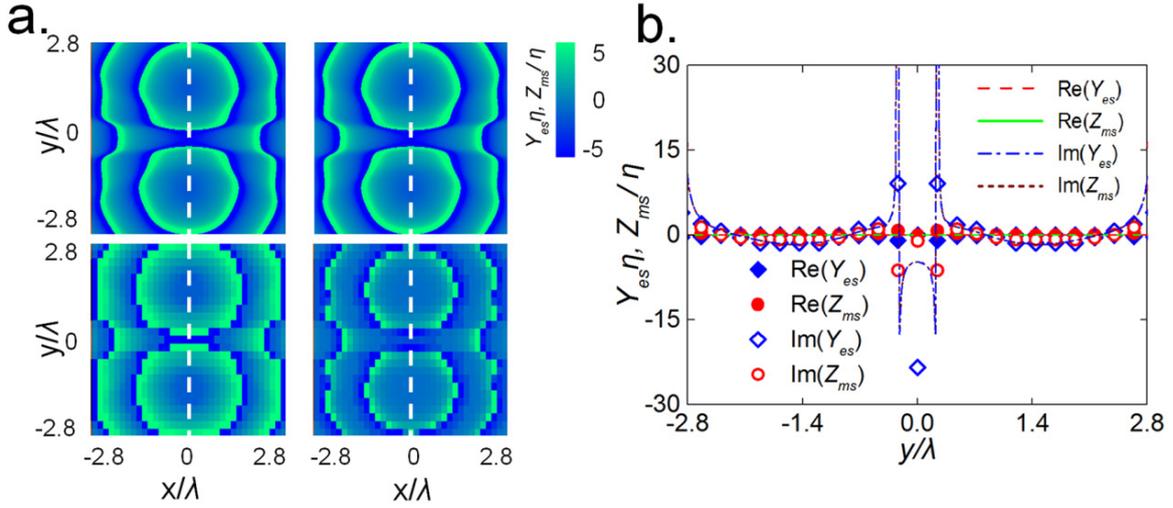

**Figure 2.** Distributions of sheet impedance. **(a)** Spatial distributions of imaginary parts of $Y_{es}\eta$ (left) and $Z_{ms}/\eta$ (right) for ideal case (upper row) as well as those realized by designed metalens (bottom row). **(b)** Distributions of real and imaginary parts of sheet impedance along dotted line at $x = 0$ in Figure 2a. Lines: ideal values. Markers: values of active Huygens' meta-atoms.

As an example of the proposed design, we have experimentally verified a 2D active metalens for arbitrary focusing in the *x-z* plane, which requires a spatially varying distribution of varactor capacitance along the *x*-direction, but invariant along the *y*-direction. For such a case, the infinite periodicity of the meta-atom array along the *y*-direction can be mimicked using perfect electric conductor (PEC) boundary conditions. Hence, to demonstrate the focusing performance of the designed metasurface, we only need to consider a linear chain of tunable meta-atoms composed of 35 elements along *x*-axis, terminated by parallel PEC plates along the *y*-direction.

The active component in the meta-atom is an off-the-shelf commercial varactor, and high resistive bias lines are added to independently and dynamically control the varactors. The metallic meta-atom is fabricated through standard printed circuit board (PCB) techniques, as shown in **Figure 3**a. All parameters are optimized to achieve adequate phase coverage with relatively high transmission amplitude, as seen in Figure 3b. The spectral phase responses are nearly linearly dependent on the bias voltage. Although the phase shift range is less than 360 degrees, it is enough for the design of the following 2D dynamic focusing demonstrations. We should emphasize that full phase coverage (360 degrees) can be achieved by a slight modification of the meta-atom design, achieved by loading more varactors into the element and making full use of electric and magnetic resonances. More details of the meta-atom design are shown in Figure S2 of the SI.

We make use of a parallel-plate waveguide near-field scanning system schematically shown in Figure 3a to experimentally explore the tunable and dynamic focusing performance of the active Huygens' metalens. A prototype of the proposed metalens composed of an array of 35 meta-atoms has been fabricated together with the voltage biasing network for each varactor, which individually control all the metalens elements. More details of the experiments are provided in the Experimental Section and the SI.

We first demonstrate the real-time dynamic focusing of a propagating EM wave, which we believe is the most significant advantage over other tunable approaches. By continuously modifying the bias-voltage profile of the meta-atoms from one distribution to another using a computer-programed multi-channel DC voltage source, we are able to scan the focal spot arbitrarily and in real time. In particular, we show that the focal spot can be dynamically controlled to move along certain routes in the *x-z* plane. Figure 3c shows the successive snap-shots of the measured electric-field amplitude distribution on the *x-z* plane. By applying a sequence of biasing-voltage profiles, recorded on the computer in advance, the focal spot (black dashed circle) is moved long a predesigned trace of the letter "N". The scanning trace can be designed in purpose by simply changing the stored sequence of biasing-voltage profiles. Experiment is also carried out to further measure the transient behavior of dynamic focusing of the reconfigurable metalens. Significantly, the control speed can be as fast as ten microseconds ($10^5$ switches per second), and details of the experiment can be found in SI.

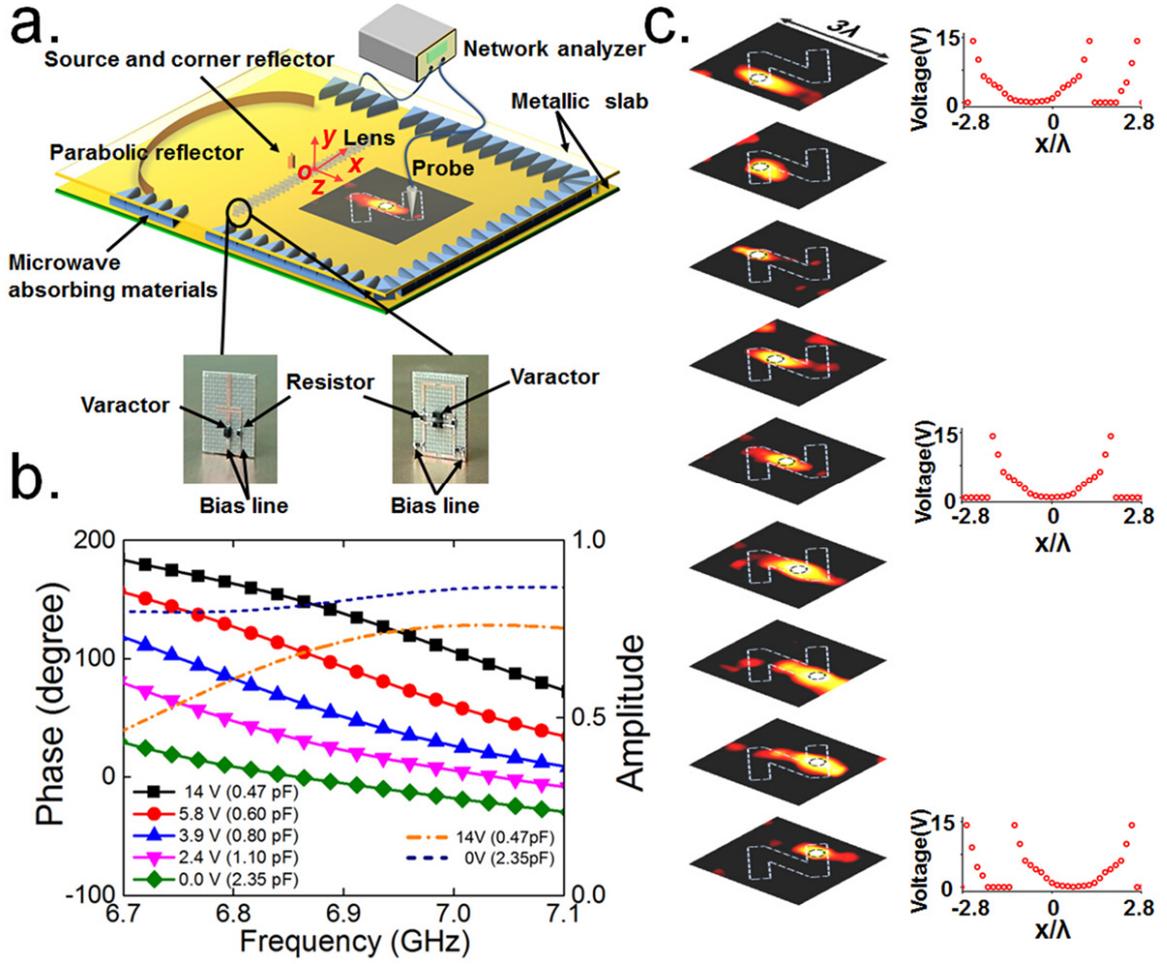

**Figure 3.** Experimental setup and measured results. **(a)** Schematic of experimental setup and photographs of frontside (right) and backside (left) of fabricated meta-atom. **(b)** Frequency-dependent transmission phase (solid-dot lines) and amplitude (dashed lines) under different voltage biasing conditions on varactors (corresponding capacitance values in parentheses). **(c)** Measured successive snapshots of electric field amplitude distribution when focal spot is dynamically moved along a trace of letter "N" (at 6.9 GHz). Black dashed circles highlighted in trace: focal spots. Right column of Figure 3c: three corresponding voltage distributions on chain of bias inputs.

To further demonstrate the flexibility and efficiency of the proposed active Huygens metalens, we show that the fabricated prototype can transform the incident wave into two focal spots, which can be controlled at will and moved to distinct positions. As shown in **Figure 4**, two foci can be arranged in front of each other along the *z*-direction (Figure 4a), or symmetrically displaced from the *z*-axis (Figure 4b), or to any other arbitrary position (Figure 4c). The corresponding distributions of bias voltages are reported in the right column of Figure 4.

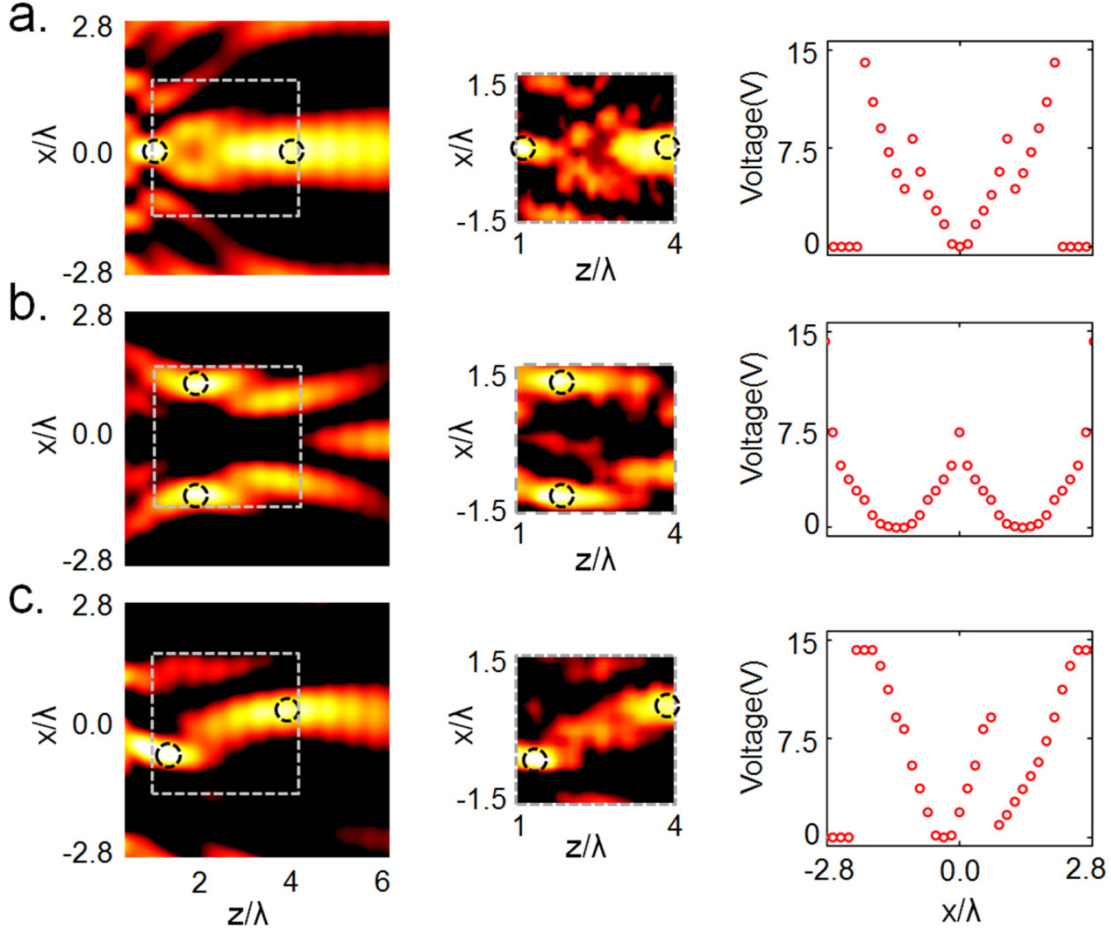

**Figure 4.** Two foci with distinct positions. Distributions of simulated (left column) and measured (middle column) amplitudes of electric field intensities ($|E_y|$) with multiple focal spots located at **(a)** $x_1 = 0$, $z_1 = \lambda$, $x_2 = 0$, $z_2 = 4\lambda$, **(b)** $x_1 = -1.3\lambda$, $z_1 = 2\lambda$, $x_2 = 1.3\lambda$, $z_2 = 2\lambda$, and **(c)** $x_1 = -0.65\lambda$, $z_1 = 1.5\lambda$, $x_2 = 0.5\lambda$, $z_2 = 4\lambda$. Right column: corresponding distributions of bias voltages.

We have also studied the focusing performance of the metalens around the operational frequency band, revealing that robust focusing can be achieved over a reasonable bandwidth, with only slightly changes in the focal distance due to the small dispersive properties of the metasurface (See Figure S9 in the SI for details). In addition, the quantitative analysis of the focal spots also indicates that the transmitted wave is well confined into the hotspots, as shown in Figure S10 of the SI. The total focusing efficiency can be defined by the fraction of power measured at the focal spot[64], normalized to the input power. In our measurement, the total focusing efficiency reaches approximately 36%. It should be noted that the modest efficiency mainly results from the loss and deviation in phase/amplitude after the introduction of realistic varactors. In fact, it is great challenging to co-design the meta-atoms with massive varactors when taking into account of their slightly varying capacitance and loss. Nevertheless, the achieved efficiency of our design is already much better than the counterparts of the existing reconfigurable metalens.[52,53] It is also reasonable to expect a more efficient transmission manipulation by using a co-design method along with low loss varactors. Besides, in all the considered cases, measurements and full-wave simulations are in good agreement. It clearly demonstrates that arbitrary single or multiple focal spots can be obtained and manipulated, in real time and with high efficiency, by tuning the bias-voltage distributions of the proposed Huygens' metalens. To achieve a higher-resolution imaging or focusing performance, one can increase the pixel number of the reconfigurable metalens. But the extra technical challenge of the biased voltage network design should be carefully considered, especially when taking consideration of the transmission efficiency.

We have explored a reconfigurable active metasurface to achieve dynamic and arbitrary controls of transmission characteristics of electromagnetic (EM) wave in real time. As an example, a reconfigurable Huygens' metalens is not only proposed for dynamical and efficient focusing of EM waves but also experimentally verified at microwave frequencies. From full-wave simulation results, the multiple foci at arbitrary positions in 3D space are verified, which can be dynamically controlled by changing the distribution of the capacitances loaded in 35×25 meta-atom arrays.This proof-of-concept implementation has been realized with 35 meta-atoms in a 2D implementation. The dynamical controls of the EM wave are experimentally verified, that can move not only the focal spot along a predesigned trajectory but also form two focal spots simultaneously and arbitarily in a 2D space by varying the programmable input voltages. Furthermore, the control speed can reach values in the order of $10^5$ switches per second. The proposed active metasurface concept may be scalable to higher frequencies, for example, in the terahertz band, where active elements may be realized by graphene or semiconductor materials.[23-26] The ability of the proposed active Huygens' metalens to arbitarily focus the incident wave in real time may find applications in smart and reconfigurable devices and systems, such as antennas and radars, sensors or imaging systems for biomedical detection and target recognition, as well as in boosting integration with other devices to create various, advanced functionalities. In general, we believe that incorporating active elements in efficient Huygens' metasurfaces holds the promise for unprecedented opportunities for device manipulation of electromagnetic fields and waves, opening untapped potential for a variety of real-world applications.

**Experimental Section**

A parallel-plate waveguide near-field scanning system (as shown in Figure 3a) is used to experimental verification of the performance of the proposed active metalens. To satisfy the periodic condition at the top and bottom sides of the array of meta-atoms shown in Figure 1a, the sample plate is inserted between two metallic plates while maintaining orthogonality with them. A monopole antenna together with a corner and a parabolic reflector is utilized to generate an incident plane wave, while another monopole antenna is used as the detector to map the electric field inside the planar waveguide utilizing a microwave vector network analyzer (Agilent N5244A).

**Supporting Information**

Supporting Information is available from the Wiley Online Library or from the author.

**Acknowledgements**

This work is partially supported by the National Nature Science Foundation of China (61671231, 61571218, 61571216, 61301017, 61371034), the Research Innovation Program for College Graduates of Jiangsu Province (KYZZ15_0028), the Welch Foundation with grant No. F-1802, the Air Force Office of Scientific Research with grant No. FA9550-13-1-0204, and partially supported by PAPD of Jiangsu Higher Education Institutions and Jiangsu Key Laboratory of Advanced Techniques for Manipulating Electromagnetic Waves. C.-W.Q. would like to acknowledge the support by the National Research Foundation, Prime Minister's Office, Singapore under its Competitive Research Program (CRP Award No. NRF-CRP15-2015-03).

**References**

[1] S. Sun, Q. He, S. Xiao, Q. Xu, X. Li, L. Zhou, *Nat. Mater.* **2012**, *11*, 426.

[2] C. Pfeiffer, A. Grbic, *Phys. Rev. Appl.* **2014**, *2*, 44012.

[3] N. Yu, P. Genevet, M. A. Kats, F. Aieta, J.-P. Tetienne, F. Capasso, Z. Gaburro, *Science* **2011**, *334*, 333.

[4] N. Yu, F. Aieta, P. Genevet, M. A. Kats, Z. Gaburro, F. Capasso, *Nano Lett.* **2012**, *12*, 6328.

[5] N. M. Estakhri, A. Alu, *Antennas Wirel. Propag. Lett. IEEE* **2014**, *13*, 1775.

[6] X. Ni, Z. J. Wong, M. Mrejen, Y. Wang, X. Zhang, *Science* **2015**, *349*, 1310.

[7] X. Ni, A. V Kildishev, V. M. Shalaev, *Nat. Commun.* **2013**, *4*, 2807.

[8] L. Huang, X. Chen, H. Mühlenbernd, H. Zhang, S. Chen, B. Bai, Q. Tan, G. Jin, K.-W. Cheah, C.-W. Qiu, J. Li, T. Zentgraf, S. Zhang, *Nat. Commun.* **2013**, *4*, 2808.

[9] D. Wen, F. Yue, G. Li, G. Zheng, K. Chan, S. Chen, M. Chen, K. F. Li, P. W. H. Wong, K. W. Cheah, *Nat. Commun.* **2015**, *6,* 8241.

[10] L. Zhang, S. Mei, K. Huang, C. Qiu, *Adv. Opt. Mater.* **2016** *4*, 818.

[11] S. B. Glybovski, S. A. Tretyakov, P. A. Belov, Y. S. Kivshar, C. R. Simovski, *Phys. Rep.* **2016**, *634*, 1.

[12] N. M. Estakhri, A. Alù, *J. Opt. Soc. Am. B* **2016**, *33*, A21.

[13] N. M. Estakhri, C. Argyropoulos, A. Alù, *Phil. Trans. R. Soc. A* **2015**, *373*, 20140351.

[14] A. Silva, F. Monticone, G. Castaldi, V. Galdi, A. Alù, N. Engheta, *Science* **2014**, *343*, 160.

[15] F. Qin, L. Ding, L. Zhang, F. Monticone, C. C. Chum, J. Deng, S. Mei, Y. Li, J. Teng, M. Hong, *Sci. Adv.* **2016**, *2*, e1501168.

[16] X. Ding, F. Monticone, K. Zhang, L. Zhang, D. Gao, S. N. Burokur, A. de Lustrac, Q. Wu, C. Qiu, A. Alù, *Adv. Mater.* **2015**, *27*, 1195.

[17] N. M. Estakhri, A. Alù, *Phys. Rev. B* **2014**, *89*, 235419.

[18] F. Monticone, N. M. Estakhri, A. Alù, *Phys. Rev. Lett.* **2013**, *110*, 203903.

[19] Y. H. Fu, A. Q. Liu, W. M. Zhu, X. M. Zhang, D. P. Tsai, J. B. Zhang, T. Mei, J. F. Tao, H. C. Guo, X. H. Zhang, *Adv. Funct. Mater.* **2011**, *21*, 3589.

[20] H.-S. Ee, R. Agarwal, *Nano Lett.* **2016**, *16*, 2818.

[21] S. Xiao, U. K. Chettiar, A. V Kildishev, V. Drachev, I. C. Khoo, V. M. Shalaev, *Appl. Phys. Lett.* **2009**, *95*, 33115.

[22] T. Driscoll, H.-T. Kim, B.-G. Chae, B.-J. Kim, Y.-W. Lee, N. M. Jokerst, S. Palit, D. R. Smith, M. Di Ventra, D. N. Basov, *Science* **2009**, *325*, 1518.

[23] W. J. Padilla, A. J. Taylor, C. Highstrete, M. Lee, R. D. Averitt, *Phys. Rev. Lett.* **2006**, *96*, 107401.

[24] H.-T. Chen, W. J. Padilla, J. M. O. Zide, A. C. Gossard, A. J. Taylor, R. D. Averitt, *Nature* **2006**, *444*, 597.

[25] S. H. Lee, M. Choi, T. T. Kim, S. Lee, M. Liu, X. Yin, H. K. Choi, S. S. Lee, C. G. Choi, S. Y. Choi, *Nat. Mater.* **2012**, *11*, 936.

[26] Z. Miao, Q. Wu, X. Li, Q. He, K. Ding, Z. An, Y. Zhang, L. Zhou, *Phys. Rev. X* **2015**, *5*, 41027.

[27] D. A. Powell, I. V Shadrivov, Y. S. Kivshar, *Appl. Phys. Lett.* **2009**, *95*, 84102.

[28] I. V Shadrivov, A. B. Kozyrev, D. W. van der Weide, Y. S. Kivshar, *Opt. Express* **2008**, *16*, 20266.

[29] D. Shrekenhamer, W. C. Chen, W. J. Padilla, *Phys. Rev. Lett.* **2013**, *110*, 278.

[30] B. Zhu, Y. Feng, J. Zhao, C. Huang, T. Jiang, *Appl. Phys. Lett.* **2010**, *97*, 51906.

[31] H. Li, D. Ye, F. Shen, B. Zhang, Y. Sun, W. Zhu, C. Li, L. Ran, *Microw. Theory Tech. IEEE Trans.* **2015**, *63*, 925.

[32] M. Liu, X. Yin, E. Ulin-Avila, B. Geng, T. Zentgraf, L. Ju, F. Wang, X. Zhang, *Nature* **2011**, *474*, 64.

[33] H.-T. Chen, W. J. Padilla, M. J. Cich, A. K. Azad, R. D. Averitt, A. J. Taylor, *Nat. Photonics* **2009**, *3*, 148.

[34] S. Zhang, J. Zhou, Y.-S. Park, J. Rho, R. Singh, S. Nam, A. K. Azad, H.-T. Chen, X. Yin, A. J. Taylor, *Nat. Commun.* **2012**, *3*, 942.

[35] G. Kenanakis, R. Zhao, N. Katsarakis, M. Kafesaki, C. M. Soukoulis, E. N. Economou, *Opt. Express* **2014**, *22*, 12149.

[36] T. J. Cui, M. Q. Qi, X. Wan, J. Zhao, Q. Cheng, *Light Sci. Appl.* **2014**, *3*, e218.

[37] N. Kaina, M. Dupré, G. Lerosey, M. Fink, *Sci. Rep.* **2014**, *4* 6693.

[38] N. Kaina, M. Dupré, M. Fink, G. Lerosey, *Opt. Express* **2014**, *22*, 18881.

[39] Y. B. Li, L. L. Li, B. B. Xu, W. Wu, R. Y. Wu, X. Wan, Q. Cheng, T. J. Cui, *Sci. Rep.* **2016**, *6* 23731.

[40] T. Sleasman, M. F. Imani, J. N. Gollub, D. R. Smith, *Appl. Phys. Lett.* **2015**, *107*, 204104.

[41] X. Wan, M. Q. Qi, T. Y. Chen, T. J. Cui, *Sci. Rep.* **2016**, *6*, 20663.

[42] H. Yang, X. Cao, F. Yang, J. Gao, S. Xu, M. Li, X. Chen, Y. Zhao, Y. Zheng, S. Li, *Sci. Rep.* **2016**, *6,* 35692.

[43] A. Pors, M. G. Nielsen, R. L. Eriksen, S. I. Bozhevolnyi, *Nano Lett.* **2013**, *13*, 829.

[44] F. Monticone, A. Alù, *Chinese Phys. B* **2014**, *23*, 47809.

[45] F. Aieta, P. Genevet, M. A. Kats, N. Yu, R. Blanchard, Z. Gaburro, F. Capasso, *Nano Lett.* **2012**, *12,* 4932.

[46] X. Chen, L. Huang, H. Mühlenbernd, G. Li, B. Bai, Q. Tan, G. Jin, C.-W. Qiu, S. Zhang, T. Zentgraf, *Nat. Commun.* **2012**, *3*, 1198.

[47] M. Khorasaninejad, W. T. Chen, R. C. Devlin, J. Oh, A. Y. Zhu, F. Capasso, *Science* **2016**, *352*, 1190.

[48] D. Lin, P. Fan, E. Hasman, M. L. Brongersma, *Science* **2014**, *345*, 298.

[49] X. Ni, S. Ishii, A. V Kildishev, V. M. Shalaev, *Light Sci. Appl.* **2013**, *2*, e72.

[50] J. Hunt, T. Driscoll, A. Mrozack, G. Lipworth, M. Reynolds, D. Brady, D. R. Smith, *Science* **2013**, *339*, 310.

[51] A. Grbic, L. Jiang, R. Merlin, *Science* **2008**, *320*, 511.

[52] W. M. Zhu, Q. H. Song, L. B. Yan, W. Zhang, P. C. Wu, L. K. Chin, H. Cai, D. P. Tsai, Z. X. Shen, T. W. Deng, *Adv. Mater.* **2015**, *27*, 4739.

[53] Q. Wang, E. T. F. Rogers, B. Gholipour, C. M. Wang, G. Yuan, J. Teng, N. I. Zheludev, *Nat. Photonics* **2016**, *10*, 60.

[54] C. Pfeiffer, A. Grbic, *Phys. Rev. Lett.* **2013**, *110,* 197401.

[55] M. Decker, I. Staude, M. Falkner, J. Dominguez, D. N. Neshev, I. Brener, T. Pertsch, Y. S. Kivshar, *Adv. Opt. Mater.* **2015**, *3*, 813.

[56] K. E. Chong, L. Wang, I. Staude, A. R. James, J. Dominguez, S. Liu, G. S. Subramania, M. Decker, D. N. Neshev, I. Brener, *ACS Photonics* **2016**, *3*, 514.

[57] M. Kim, A. M. H. Wong, G. V Eleftheriades, *Phys. Rev. X* **2014**, *4*, 41042.

[58] A. Epstein, G. V Eleftheriades, *J. Opt. Soc. Am. B* **2016**, *33*, A31.

[59]  A. Epstein, J. P. S. Wong, G. V Eleftheriades, *Nat. Commun.* **2016**, *7*, 10360.

[60]  B. O. Zhu, K. Chen, N. Jia, L. Sun, J. Zhao, T. Jiang, Y. Feng, *Sci. Rep.* **2014**, *4*.

[61]  W. L. Stutzman, G. A. Thiele, *Antenna Theory and Design*, John Wiley & Sons, **2012**.

[62]  W. Wang, Z. Guo, K. Zhou, Y. Sun, F. Shen, Y. Li, S. Qu, S. Liu, *Opt. Express* **2015**, *23*, 29855.

[63]  R. Li, Z. Guo, W. Wang, J. Zhang, K. Zhou, J. Liu, S. Qu, S. Liu, J. Gao, *Photonics Res.* **2015**, *3*, 252.

[64]  A. Arbabi, Y. Horie, A. J. Ball, M. Bagheri, A. Faraon, *Nat. Commun.* **2015**, *6*, 7069.